\documentclass[10pt,a4paper]{article}
\usepackage{epsfig,amssymb,amsmath}
\begin{document} 
\title{\vskip -70pt \begin{flushright} {\normalsize
DAMTP-2006-84} \\ \end{flushright} 
\vskip 80pt 
{Reparametrising the Skyrme Model using the Lithium-6 Nucleus}
\vskip 20pt} 
\author{Nicholas S.
Manton\thanks{N.S.Manton@damtp.cam.ac.uk}\,\,\,and\,\,\,Stephen W.
Wood\thanks{S.W.Wood@damtp.cam.ac.uk} \\ \\DAMTP, Centre for
Mathematical Sciences,\\University of Cambridge, Wilberforce Road,\\ 
Cambridge CB3 0WA, UK}
\maketitle 
\begin{abstract} 
The minimal energy $B=6$ solution of the Skyrme model is a static
soliton with $D_{4d}$ symmetry. The symmetries of the solution imply 
that the quantum numbers of the ground state are the same as
those of the Lithium-6 nucleus. This identification is considered
further by obtaining expressions for the mean charge radius and
quadrupole moment, dependent only on the Skyrme model parameters $e$
(a dimensionless constant) and $F_\pi$ (the pion decay constant). The
optimal values of these parameters have often been deliberated
upon, and we propose, for $B>2$, changing them from those which are
most commonly accepted. We obtain specific
values for these parameters for $B=6$, by matching with properties of
the Lithium-6 nucleus. We find further support for the new values by
reconsidering the $\alpha$-particle and deuteron as quantized $B=4$
and $B=2$ Skyrmions.
\end{abstract}

\section{Introduction}
A nonlinear theory was put forward by T. H. R. Skyrme in 1961
\cite{skyrme}. It sets out to describe a low-energy effective theory
of QCD, and the Lagrangian of the theory is defined in terms of pion
fields. The most successful and promising physical interpretation of
the theory is as a description of atomic nuclei. The Skyrme model
admits topological soliton solutions, falling into sectors labelled by
an integer-valued topological degree. This quantity is identified with
baryon number. A quantized Skyrmion of topological charge $B$ is
interpreted as a nucleus with baryon number $B$.

To date, relatively little work has been done on the electromagnetic
properties of quantized Skyrmions and their comparison to known
experimental results. Two papers by Braaten and Carson
\cite{bc,bc1} consider the electrostatic properties of the $B=2$
Skyrmion, and their comparison to the experimentally determined
properties of the deuteron. This analysis was extended in \cite{lms}
to account for the softest vibrational modes of the $B=2$ Skyrmion, the
quantization of which resulted in an accurate prediction of the mean
charge radius of the deuteron.

In this paper we perform a similar analysis for the $B=6$
Skyrmion. However, performing a quantization of all possible
vibrational modes for $B>2$ is technically very difficult. We
therefore consider rescaling the Skyrme model parameters (introduced
in the next section), whilst suppressing the vibrational modes, in
order to calculate observables that more closely correspond to 
experiment.

The outline of the paper is as follows. In the following section we
present the Skyrme model and, following \cite{callanwitten}, give an
expression for the electromagnetic current. Section \ref{sec:rma}
provides an overview of the rational map ansatz for Skyrmions 
\cite{houghton}, which
is used in later sections. Section \ref{sec:ferm} discusses the fermionic
quantization of solitons, as applied to the Skyrme model, recalls the
Finkelstein-Rubinstein (FR) constraints \cite{fr}, and reviews
Krusch's method of determining the FR constraints \cite{krusch}. In
section \ref{sec:zero} we recall the semi-classical method of 
soliton quantization, following \cite{bc}, in which only the
translational, rotational, and isorotational collective coordinates
are considered. 
In section \ref{sec:quant} this quantization of the $B=6$
Skyrmion is performed, and we find that the symmetries of the minimal-energy
classical solution, via the FR constraints, ensure that the
lowest-lying quantized state has spin 1 and isospin 0, 
the quantum numbers of the Lithium-6
nucleus. In section \ref{sec:rad} we present an expression for the
mean charge radius of the quantized $B=6$ Skyrmion. Section
\ref{sec:quad} presents an expression for the quadrupole moment of the
quantized $B=6$ Skyrmion. The reparametrisation of the Skyrme model in
the $B=6$ sector is considered in section \ref{sec:reparam}, and the
mean charge radius and quadrupole moment of the quantized $B=6$
Skyrmion are explicitly calculated and compared to experiment. In
section \ref{sec:b=2} we reconsider the $B=2$ and $B=4$ Skyrmions in 
the light of the reparametrisation. We provide a
conclusion in section \ref{sec:conc}. The calculations presented in
this paper support our interpretation of the Lithium-6 nucleus as the
lowest-lying quantum state of the $B=6$ Skyrmion. They confirm
that if one wishes to apply the Skyrme model to nuclei, then a
reparametrisation is desirable.

\section{The Skyrme Model}

The Skyrme model is a nonlinear theory of pions in three spatial
dimensions, defined in terms of an $SU(2)$-valued scalar, the Skyrme
field \cite{skyrme,manton}. It provides a low energy effective theory of QCD,
becoming exact as the number of quark colours becomes large
\cite{wittenglobal,wittencurrent}. Its topological soliton solutions,
Skyrmions, can be interpreted as nucleons and nuclei.

The $SU(2)$ Skyrme model has the Lagrangian
density 
\begin{equation}
\label{eq:lag}
\mathcal{L} = \frac{F_\pi^2}{16}\,\hbox{Tr}\,\partial_\mu U
\partial^\mu U^{-1} + \frac{1}{32e^2}\,\hbox{Tr}\,[\partial_\mu U U^{-1},
\partial_\nu U U^{-1}][\partial^\mu U U^{-1},
\partial^\nu U U^{-1}] + \frac{1}{8} m_\pi ^2 F_\pi
^2\,\hbox{Tr}\,(U-1_2) \,,
\end{equation}
where $U(t,\mathbf{x})$ is the Skyrme field, $F_\pi$ is the pion 
decay constant, $e$ is a
dimensionless parameter and $m_\pi$ is the pion mass.

It is helpful to use energy and length units of $F_\pi / 4e$ and
$2/eF_\pi$ respectively. In terms of these Skyrme units the Skyrme 
Lagrangian becomes
\begin{equation}
\label{eq:lagstandard}
L=\int \left\{ -\frac{1}{2}\,\hbox{Tr}\,(R_\mu R^\mu) 
+ \frac{1}{16}\,\hbox{Tr}\,([R_\mu,R_\nu][R^\mu,R^\nu]) 
+ m^2\,\hbox{Tr}\,(U - 1_2) \right\} d^3 x \,,
\end{equation}
where we have introduced the $\mathfrak{su}(2)$-valued current $R_\mu
= (\partial_\mu U)U^{-1}$, and defined $m = 2m_\pi / eF_\pi$.

The scalar field $U$, at a fixed time, is a map
from $\mathbb R ^3$ into $S^3$, the group manifold of $SU(2)$. However, the
boundary condition $U \rightarrow 1_2$ at spatial infinity implies a
one-point compactification of space, so that topologically $U: S^3
\rightarrow S^3$, where the domain $S^3$ is identified with $\mathbb R ^3\cup
\{ \infty \}$. 
Thus, configurations $U$ obeying this boundary condition fall
into topological sectors labelled by their topological degree 
\begin{equation} 
B = \int B_0(\mathbf{x}) \, d^3 x
\end{equation} 
where 
\begin{equation}
\label{eq:cur}B_\mu(x)=\frac{1}{24\pi^2}\,
\epsilon_{\mu\nu\alpha\beta}\,\hbox{Tr}\, \partial^\nu
UU^{-1}\partial^\alpha U U^{-1}\partial^\beta U U^{-1} \,.
\end{equation}
The degree $B$, which takes integer values, is identified with baryon 
number. We refer to $B_0$
as the baryon density. The minimal energy, static solutions of the
field equation for each nonzero baryon number we call Skyrmions.

The internal symmetry of the Skyrme model is the global isospin
symmetry $U \rightarrow AUA^\dag$ where $A$ is an $SU(2)$ matrix. This
is generated by the infinitesimal transformations $U \rightarrow
U + i \epsilon [\tau_p , U]$, where $\tau_p : p=1,2,3$ are the Pauli
matrices. To couple electromagnetism to the Skyrme model, the $U(1)$
symmetry generated by the third component of isospin is gauged, with the
derivatives in (\ref{eq:lag}) and (\ref{eq:lagstandard}) replaced by 
the covariant derivative
$D_\mu U= \partial_\mu U - ie_0A_\mu[Q,U]$,
where $Q = \frac{1}{2}\tau_3 + \frac{1}{6}1_2$ is the charge matrix of quarks.
In the presence of electromagnetism,
the form of the baryon current density given previously is unsatisfactory
as it is not gauge invariant under 
\begin{equation}
U(x)\rightarrow U(x) +
ie_0\alpha(x)[Q,U] \,, \,\,\,A_\mu\rightarrow A_\mu +
\partial_\mu\alpha \,.
\end{equation}
The gauge invariant, conserved generalisation is given by \cite{callanwitten}
\begin{eqnarray}
B_\mu (x)&=& \frac{1}{24\pi^2}\epsilon_{\mu\nu\alpha\beta}
\Big\{\,\hbox{Tr}\,U^{-1}\partial^\nu
UU^{-1}\partial^\alpha U U^{-1}\partial^\beta U + 3ie_0A^\nu \,
\hbox{Tr}\,Q(U^{-1}\partial^\alpha U U^{-1}\partial^{\beta}U \nonumber \\
&&\quad - \partial^\alpha U
U^{-1}\partial^\beta U U^{-1}) + 3ie_0\partial^\nu A^\alpha
\,\hbox{Tr}\,Q(U^{-1}\partial^\beta U 
+ \partial^\beta U U^{-1})\Big\} \,. 
\end{eqnarray} 
The electromagnetic current is determined as 
\begin{eqnarray} 
J_\mu(x)=\frac{1}{16\pi^2}\epsilon_{\mu\nu\alpha\beta}\,\hbox{Tr}\,Q(\partial^\nu U U^{-1}
\partial^\alpha U U^{-1} \partial^\beta U U^{-1} + U^{-1}\partial^\nu U
U^{-1} \partial^\alpha U U^{-1} \partial^\beta U) \nonumber\\
+ \frac{ie_0}{4\pi^2}\epsilon_{\mu\nu\alpha\beta}\partial^\nu
A^\alpha\,\hbox{Tr}\,\Big\{Q^2\partial^\beta U U^{-1} 
+ Q^2U^{-1}\partial^\beta U + \frac{1}{2}Q\partial^\beta UQU^{-1} - \frac{1}{2}QUQ\partial^\beta U^{-1}\Big\} +
I_\mu^3(x) \,. 
\end{eqnarray}

We are interested in static or slowly varying Skyrme fields, quantized
to have isospin zero. In this case the electromagnetic current
simplifies. It can be shown, assuming 
$I_0^3(x)=0$, $A_0 \neq 0$, $\hbox{\bf A}=0$
and $\partial_0 A_\mu = 0,$ that 
\begin{equation} 
\label{eq:jb}
J_0 = \frac{1}{2}B_0 = \frac{1}{24\pi^2}\epsilon_{0ijk}\epsilon_{pqr}\gamma_i^p\gamma_j^q\gamma_k^r =
\frac{1}{4\pi^2} \det \Gamma \,, 
\end{equation} 
where the matrix $\Gamma$ has elements $\gamma_i^p$ defined by 
\begin{equation} 
(\partial_i U)U^{-1} = -i\gamma_i^p \tau_p \,. 
\end{equation} 
Therefore, the electric charge density is half the original baryon density.

\section{The Rational Map Ansatz}
\label{sec:rma} 
The rational map ansatz \cite{houghton} applies the topological notion
of suspension, using a rational map from $S^2 \rightarrow S^2$ to
construct an approximate Skyrmion, a map from $\mathbb R ^3 \rightarrow
S^3$. It exhibits a nonlinear separation of variables,
separating the angular and radial dependence of the Skyrme 
field. One identifies the domain $S^2$ with concentric spheres in
$\mathbb R ^3$, and the target $S^2$ with spheres of latitude on
$S^3$. A point in $\mathbb R ^3$ can be denoted by its coordinates
$(r,z)$, where $r$ is the radial distance from the origin and $z$
specifies the direction from the origin, a point on the unit
sphere. Via stereographic projection, the complex coordinate $z$ can
be identified with conventional polar coordinates by 
$z = \hbox{tan}(\theta /2)e^{i\phi}$. 
Equivalently, the point $z$ corresponds to the unit vector
\begin{equation} 
\mathbf{n}_z = \frac{1}{{1 + |z|^2}}(z + \bar{z},i(\bar{z}-z), 1 -
| z|^2) \,, 
\end{equation} 
and conversely
\begin{equation}
z = \frac{(\mathbf{n}_z)_1 
+ i(\mathbf{n}_z)_2}{1+(\mathbf{n}_z)_3} \,.
\end{equation}

The ansatz for the Skyrme field depends on a rational map 
$R(z) = p(z) / q(z)$, where $p$
and $q$ are polynomials in $z$, and a radial profile function
$f(r)$. The value of the rational map $R$ is associated with the unit
vector 
\begin{equation} 
\mathbf{n}_R = \frac{1}{{1 + |R|^2}}(R +
\bar{R},i(\bar{R}-R), 1 - |R|^2) \,. 
\end{equation} 
The ansatz is 
\begin{equation}
\label{eq:ansatz} 
U(r,z) = \hbox{exp}\,(if(r)\mathbf{n}_{R(z)} \cdot
\boldsymbol{\tau}) \,, 
\end{equation} 
with $f(r)$ satisfying $f(0)=\pi$ and $f(\infty)=0$. 

The degree of the rational map, $N$, is the greater of the algebraic degrees
of the polynomials $p$ and $q$. $N$ is also equal to the topological degree
of the map $R$ viewed as a map from $S^2 \rightarrow S^2$. The baryon
density of the ansatz is given by 
\begin{equation} 
\label{eq:baryon}
B_0(\mathbf{x}) = 
\frac{-f'}{2\pi^2}{\left({\frac{\hbox{sin }f}{r}}\right)}^2
\left( \frac{1 + | z|^2}{1 + |R|^2} \left\vert \frac{dR}{dz}\right\vert
\right)^2 \,, 
\end{equation}
and so the baryon number is given by
\begin{equation}
B = \int \frac{-f'}{2\pi^2}{\left({\frac{\hbox{sin }f}{r}}\right)}^2\left( \frac{1 +
| z|^2 }{ 1 + |R|^2} \left\vert\frac{dR }{ dz }\right\vert
\right)^2\,\frac{2i\,dz\,d\bar{z}}{(1+|z|^2)^2}\,r^2\,dr \,, 
\end{equation}
where $2i\,dz\,d\bar{z}/(1+|z|^2)^2$ is equivalent to the usual 2-sphere
area element $\hbox{sin}\,\theta\,d\theta\,d\phi$. It is straightforward to
show that $B=N$.

An $SU(2)$ M\"{o}bius transformation $M_1$ of $R$ corresponds to an isorotation; an $SU(2)$ M\"{o}bius transformation $M_2$ of $z$ 
corresponds to a rotation in physical space. They induce the following 
transformation of the rational map $R(z)$: 
\begin{equation} 
\label{eq:symm}
R(z) \rightarrow \tilde{R}(z)=M_1(R(M_2(z))) \,. 
\end{equation} 

The energy for a field of the form (\ref{eq:ansatz}) is 
\begin{eqnarray} 
E= \int \Bigg\{\, f'^2 + 2\frac{\sin^2 f}{r^2}(f'^2+1) 
\left(\frac{1+|z|^2}{1+|R|^2}\left\vert\frac{dR
}{ dz }\right\vert \right)^2 + \frac{\sin^4 f}{r^4} \left(
\frac{1+|z|^2}{1+|R|^2}\left\vert\frac{dR }{dz }\right\vert \right)^4 
\nonumber \\
+ 2m^2(1-\cos f)\,\Bigg\}\,
\frac{2i\,dz\,d\bar{z}}{(1+|z|^2)^2}\,r^2\,dr \,,
\end{eqnarray} 
which can be simplified to 
\begin{equation} 
\label{eq:e}
E=4\pi \int_{0}^{\infty} \left(r^2 f'^2 + 2B\sin^2
f(f'^2+1)+ \mathcal{I}\,\frac{\sin^4 f}{r^2} + 2m^2r^2(1-\cos
f)\right) dr \,. 
\end{equation}
Here $\mathcal{I}$ denotes the purely angular integral 
\begin{equation} 
\mathcal{I} = \frac{1}{4\pi} \int
\left(\frac{1+|z|^2}{1+|R|^2}\left\vert\frac{dR }{ dz }\right\vert
\right)^4\frac{2i\,dz\,d\bar{z}}{(1+|z|^2)^2} 
\end{equation} 
which only depends on the rational map $R$. Both $B$ and
$\mathcal{I}$, and hence the energy $E$, are invariant under the
transformations (\ref{eq:symm}).

To minimise $E$ one first minimises $\mathcal{I}$ over all maps of
degree $B$. The profile function $f$ may then be
found by solving the second order ODE that is the Euler-Lagrange
equation for the expression (\ref{eq:e}) with $B$ and $\mathcal{I}$ as fixed
parameters:
\begin{equation}
(r^2 + 2B\sin^2 f)f'' + 2rf' + \sin 2f \left(B(f'^2 - 1) 
- \mathcal{I}\,\frac{\sin^2 f}{r^2}\right) - m^2 r^2 \sin f = 0 \,.
\end{equation}
Note that the inclusion of the pion mass term in the
Lagrangian density has no effect on the rational map, but results in
the profile function being slightly modified, leading to higher
energies than in the massless case. The optimised rational map ansatz
has been shown to give a good approximation to the true Skyrmion for
baryon numbers up to $B=7$ (and for a larger range of $B$ in the
massless pion case) \cite{bs1,houghton}. For $B=8,9$ and beyond, and
$m$ of order 1, the structure of the minimal energy Skyrmions differs
qualitatively from that given by the rational map ansatz \cite{bms,hm}.

\section{Fermionic Quantization of the Skyrme Model}
\label{sec:ferm} 
We recall that the configuration space $\cal{C}$ of the Skyrme model
has connected components ${\cal{C}}_B$, which are the homotopy
classes, labelled by baryon number. The fundamental group of
each component satisfies $\pi_1({\cal{C}}_B)=\mathbb Z _2$, and
therefore all ${\cal{C}}_B$
admit double-valued functions. However, these double-valued functions can
be defined as single-valued functions on $\widetilde{{\cal{C}}_B}$,
the universal covering space of ${\cal{C}}_B$.

We write $\tilde{q},\tilde{q}'$ for different points of
$\widetilde{{\cal{C}}_B}$ covering the point 
$q \in {\cal{C}}_B$. The condition that $\tilde{q} \neq \tilde{q}'$ 
implies that a path from $\tilde{q}$ to
$\tilde{q}'$ projects to a non-contractible loop in
${\cal{C}}_B$. Double-valued functions $\Psi$ on ${\cal{C}}_B$ are
thought of as functions on $\widetilde{{\cal{C}}_B}$: 
\begin{equation} 
\Psi:\widetilde{{\cal{C}}_B} \rightarrow \mathbb
C,\,\,\,\Psi=\Psi(\tilde{q}) \,. 
\end{equation} 
Fermionic quantization requires that the wavefunction is defined 
on $\widetilde{{\cal{C}}_B}$ and satisfies
\begin{equation}
\Psi(\tilde{q})=-\Psi(\tilde{q}') \,.
\end{equation}

Let a rational map $R$ have the following symmetry, for some
particular $M_1$ and $M_2$, 
\begin{equation}
R(z)=M_1(R(M_2(z))) \,, 
\end{equation} 
where $M_1$ corresponds to an isorotation by
$\theta_1$ around $\mathbf{n}_1$ and $M_2$ corresponds to a rotation by $\theta_2$ around $\mathbf{n}_2$. For $\theta_2 \neq 2\pi k$
for $k \in \mathbb Z$, $M_2$ only leaves the antipodal points 
\begin{equation} 
z_{\mathbf{n}_2} =
\frac{(\mathbf{n}_2)_1 + i(\mathbf{n}_2)_2}{1 +
(\mathbf{n}_2)_3}\,\,\,\hbox{and}\,\,\,z_{-\mathbf{n}_2} =
-\frac{(\mathbf{n}_2)_1 + i(\mathbf{n}_2)_2}{1 - (\mathbf{n}_2)_3} 
\end{equation}
fixed. Similarly, $M_1$ only leaves $R_{\pm \mathbf{n}_1}$ fixed,
where $R_{\pm \mathbf{n}_1}$ are defined similarly as above. Therefore, for the
symmetry $R(z)=M_1(R(M_2(z)))$ to hold, we have 
\begin{equation} 
R(z_{-\mathbf{n}_2}) = R_{\pm \mathbf{n}_1} \,. 
\end{equation} 
By reversing the signs of $\mathbf{n}_1$ and $\theta_1$ if necessary, 
we may set 
\begin{equation} 
R(z_{-\mathbf{n}_2}) = R_{\mathbf{n}_1} \,. 
\end{equation}

Now consider the Skyrme field configuration defined by the rational
map ansatz, using this rational map and some profile function $f$, and
assume the quantum wavefunction is non-vanishing for this field
configuration. The symmetry gives rise to a loop in configuration
space (one thinks of this as a loop in ${\cal{C}}_B$ by letting the isorotation angle increase from 0 to $\theta_1$, with the rotation
angle increasing from 0 to $\theta_2$). This leads to the following
constraint on the wavefunction: 
\begin{equation}
e^{i\theta_2\mathbf{n}_2\cdot\mathbf{L}}
e^{i\theta_1\mathbf{n}_1\cdot\mathbf{K}}|\Psi\rangle
= \chi_{FR}|\Psi\rangle \,, 
\end{equation} 
where $\mathbf{L}$ and $\mathbf{K}$ are the body-fixed
spin and isospin operators respectively, and the Finkelstein-Rubinstein (FR)
phase $\chi_{FR}$ enforces the fermionic quantization condition:
\begin{equation} 
\chi_{FR} = \left\{ \begin{array}{ll} +1 & \textrm{if
the loop induced by the symmetry is contractible,} \\ -1 &
\textrm{otherwise.} \end{array} \right. 
\end{equation}

We note the following general result, proved in \cite{krusch}:

The value of $\chi_{FR}$ for a given symmetry of a rational map only
depends on the isorotation angle $\theta_1$ and the rotation angle
$\theta_2$, where the angles are defined such that $R(z_{-\mathbf{n}_2}) =
R_{\mathbf{n}_1}$, and is given by 
\begin{equation} 
\chi_{FR} = (-1)^{\cal{N}} \,,\,\,\,\hbox{where}
\,\,\,{\cal{N}}=\frac{B}{2\pi}(B\theta_2 - \theta_1) \,. 
\end{equation}

\section{Semi-Classical Collective Coordinate Quantization} 
\label{sec:zero}
The Skyrme Lagrangian is invariant under the Poincar\'{e} group of
$(3+1)$-dimensional space, $SO(3)$ isorotations and some discrete
parity transformations, which will not be 
considered here. Similarly, the space of
static solutions, that is, configurations which minimise the energy
functional, is invariant under the Euclidean group and isorotations,
$\mathbb E_3 \times SO(3)^I$. By acting with the latter symmetry group
on a static Skyrmion $U_0$ we generate a set of new 
solutions:
\begin{equation} 
U(\mathbf{x})=A_1U_0(D(A_2)(\hbox{\bf x}-\mathbf{X}))A_1^{\dag} \,,
\end{equation} 
where $A_1, A_2$ are $SU(2)$ matrices and $D(A_2)_{ij} = \frac{1}{2}
\hbox{Tr}(\tau_iA_2\tau_j A_2^{\dag})$ is the associated $SO(3)$
rotation. The classical degeneracy of the above solutions is removed 
when the theory is quantized. We think of the parameters
$\mathbf{X}(t)$, $A_1(t)$ and $A_2(t)$ as dynamical variables and then quantize
the resulting dynamical system according to standard canonical methods
\cite{bc}. This is the collective coordinate quantization of the
Skyrmion $U_0$. As
we shall only be concerned with the computation of spin and isospin, we
shall ignore the translational degrees of freedom and quantize the solitons
in their zero-momentum frame. We shall also ignore possible
deformations of the Skyrmion $U_0$, which lead to 
vibrational excitations.

Our dynamical ansatz is then $\hat{U}(\hbox{\bf
x},t)=A_1(t)U_0(D(A_2(t))\hbox{\bf x})A_1(t)^{\dag}$. Inserting this into the
Skyrme Lagrangian, one obtains the kinetic contribution to the total
energy as 
\begin{equation} 
T=\frac{1}{2}a_i U_{ij} a_j - a_i W_{ij} b_j + \frac{1}{2}b_i V_{ij}
b_j \,, 
\end{equation} 
where
\begin{equation} 
a_j = -i\,\hbox{Tr}\,\tau_j {A_1}^{\dag}\dot{A}_1,\,\,\,b_j =
i\,\hbox{Tr}\,\tau_j \dot{A}_2{A_2}^{\dag} \,, 
\end{equation} 
and the inertia tensors $U_{ij}$,
$W_{ij}$ and $V_{ij}$, expressed as functionals of $U_0(\mathbf{x})$, are
given by \cite{krusch}:
\begin{eqnarray}
U_{ij} &=& -\int \hbox{Tr}\,
\left(T_iT_j + \frac{1}{4}[R_k,T_i][R_k,T_j]\right)
\, d^3 x \,,\\
W_{ij} &=& \int \epsilon_{jlm}\,x_l\,\hbox{Tr}\,
\left(T_iR_m + \frac{1}{4}[R_k,T_i][R_k,R_m]\right) \, d^3 x \,,\\
V_{ij} &=& -\int \epsilon_{ilm}\,\epsilon_{jnp}\,x_lx_n\,
\hbox{Tr}\,\left(R_mR_p + \frac{1}{4}[R_k,R_m][R_k,R_p]\right) \, d^3 x \,,
\end{eqnarray} 
where $R_k = (\partial_k U_0)U_0^{-1}$ is the right invariant 
$\mathfrak{su}(2)$ current defined previously and
\begin{equation}
T_i = \frac{i}{2}\left[\tau_i,U_0\right]U_0^{-1}
\end{equation}
is also an $\mathfrak{su}(2)$ current. The potential energy, in terms
of collective coordinates, is just a constant, the static mass of the Skyrmion.

The quantized momenta corresponding to $b_i$ and $a_i$ become the
body-fixed spin and isospin angular momenta $L_i$ and $K_i$ satisfying
the $\mathfrak{su}(2)$ commutation relations \cite{bc}. The usual
space-fixed spin and isospin angular momenta $J_i$ and $I_i$ are
related to the body-fixed operators by
\begin{equation} 
J_i=-D(A_2)_{ij}^T L_j,\,\,\,I_i = -D(A_1)_{ij}K_j \,. 
\end{equation}
We also have $\mathbf{J}^2 = \mathbf{L}^2$, $\mathbf{I}^2 = \mathbf{K}^2$.
Thus the operators $\mathbf{J}$, $\mathbf{L}$, $\mathbf{I}$ and
$\mathbf{K}$ form the Lie algebra of $SO(4)_{J,L} \otimes SO(4)_{I,K}$. Their
action on $A_1$ and $A_2$ is given by 
\begin{equation}
[J_i,A_2]=\frac{1}{2} A_2 \tau_i\,,\,\,\,[I_i,A_1]=-\frac{1}{2}\tau_i
A_1 \,, 
\end{equation} 
\begin{equation}
[L_i,A_2]=-\frac{1}{2}\tau_i A_2 \,,\,\,\,[K_i,A_1]=\frac{1}{2} A_1 \tau_i
 \,, 
\end{equation} 
with all other commutators between momenta and coordinates zero.

A basis for the Hilbert space of states is given by $|J,J_3,L_3\rangle
\otimes |I,I_3,K_3\rangle$, with $-J \leq J_3$, $L_3 \leq J$ and $-I \leq
I_3$, $K_3 \leq I$. Concretely, $|J,J_3,L_3\rangle$ and
$|I,I_3,K_3\rangle$ are Wigner functions of the Euler angles
parametrising $A_2$ and $A_1$ respectively. The ground states are the
states with the lowest values of $J$ and $I$ that are compatible with
the FR constraints arising from the symmetries of the given
Skyrmion. The allowed values of $L_3$ and $K_3$ are also constrained
by the symmetries of the Skyrmion. In what follows, the arbitrary
third components of the space and isospace angular momenta $J_3$ and
$I_3$ are omitted.

Recall that physical states satisfy \cite{fr} 
\begin{equation} 
e^{2\pi i\mathbf{n}\cdot\mathbf{L}}|\Psi\rangle 
= e^{2\pi i\mathbf{n}\cdot\mathbf{K}}|\Psi\rangle 
= (-1)^B|\Psi\rangle \,, 
\end{equation} 
so even $B$ implies that the spin and isospin,
$J$ and $I$, are integral, and odd $B$ implies that they are
half-integral.

For general $B>1$, the moments of inertia are larger for rotations 
than for isorotations. Since these appear in the denominator of the
quantum Hamiltonian, the quantum states of lowest energy are those 
with minimal isospin, and spin excitations of Skyrmions require less energy
than isospin excitations. In particular, for even $B$ the ground
state has zero isospin, but (because of the FR constraints) not
necessarily zero spin. These observations match what is seen
experimentally for a large range of nuclei up to baryon number about 40.

\section{Quantization of the $B=6$ Skyrmion} 
\label{sec:quant}
The minimal energy $B=6$ Skyrmion has $D_{4d}$ symmetry, and is well
approximated by the rational map ansatz. The optimised rational map, in a
convenient orientation, is \cite{houghton}
\begin{equation}
\label{eq:ratmap}
R(z)=\frac{z^4 + a }{ z^2 (az^4 + 1)},\,\,\,a=0.16i \,.
\end{equation}
The $D_{4d}$ symmetry can be seen by considering the expression for
the baryon density in (\ref{eq:baryon}). The baryon density vanishes
where the Wronskian $W$ of the map $R = p/q$, 
given by
\begin{equation}
W = qp'-pq' = -2z(az^8+(3a^2-1)z^4+a) \,,
\end{equation}
vanishes. In addition to the nine zeros of $W$, the baryon density
also vanishes at $z=\infty$ due to the factor of $z^2$ in the
denominator of the rational map. The polyhedron associated with the
$B=6$ Skyrmion consists of two halves, each formed from a square with
pentagons attached to all four sides. To join these two halves, the
two squares must be parallel, with the bottom one rotated by $\pi/4$
relative to the top one. The ten face centres of the
polyhedron are the zeros of the baryon density, and the edges and
vertices of the polyhedron is where the baryon density is 
concentrated.

The $D_{4}$ subgroup is generated by two elements, a $\pi$ rotation about
the $x_1$-axis, and a $\pi$ rotation about the $x_1 = x_2$ axis. The
product of these is a $\pi /2$ rotation about the $x_3$-axis.
The corresponding symmetries of the rational map are $R(z)=1/R(1/z)$ and
$R(z)=-1/R(i/z)$. Therefore, the solution is invariant under the following
two symmetries:

$\bullet$ \,\,\,a $\pi$ rotation about the $x_1$-axis combined with a $\pi$
isorotation about the 1-axis, and

$\bullet$ \,\,\,a $\pi$ rotation about the $x_1 = x_2$ axis combined with a
$\pi$ isorotation about the 2-axis.

For the first symmetry, we have $\mathbf{n}_1 = \mathbf{n}_2 = (1,0,0)^T$,
and $\theta_1 = \theta_2 = \pi$. As required, $R(z_{-\mathbf{n}_2})
= R_{\mathbf{n}_1} = 1$. We compute ${\cal{N}}=15$ using the formula from the
previous section, and deduce that $\chi_{FR}=-1$. For the second symmetry, 
we have $\mathbf{n}_1 = (0,1,0)^T$, $\mathbf{n}_2 =
\frac{1}{\sqrt{2}}(1,1,0)^T$, and $\theta_1 =
\theta_2 = \pi$. As required we have $R(z_{-\mathbf{n}_2}) =
R_{\mathbf{n}_1} = i$. Again ${\cal{N}}=15$, so $\chi_{FR}=-1$. 
Therefore the FR constraints reduce to
\begin{equation}
e^{i\pi L_1}e^{i\pi K_1}|\Psi\rangle = -|\Psi\rangle \,,
\end{equation}
and
\begin{equation}
e^{i\pi(L_1 + L_2)/\sqrt{2}}e^{i\pi K_2}|\Psi\rangle = -|\Psi\rangle \,.
\end{equation}
We note that these constraints are equivalent to those obtained in
\cite{irwin} using an alternative method by which the contractibilities of
the closed loops corresponding to the symmetry group elements were
determined by continuous deformation of the minimal energy solution into
three well-separated $B=2$ tori.

A basis for the Hilbert space of states is given by $|J,L_3\rangle \otimes
| I,K_3\rangle$. We require the identities \cite{landau}: 
\begin{eqnarray} 
e^{i\pi L_1} |J,L_3\rangle &=&
\sum_{L_3'} |J,L_3'\rangle D_{L_3'L_3}^J (0,\pi,\pi) = (-1)^J
| J,-L_3\rangle \,, \\
e^{i\pi K_1} |I,K_3\rangle &=& \sum_{K_3'}
| I,K_3'\rangle D_{K_3'K_3}^I (0,\pi,\pi) = (-1)^I |I,-K_3\rangle \,, \\
e^{i\pi K_2} |I,K_3\rangle &=& \sum_{K_3'} |I,K_3'\rangle D_{K_3'K_3}^I
(0,\pi,0) = (-1)^{I+K_3} |I,-K_3\rangle \,, \\ 
e^{i\pi(L_1 + L_2)/\sqrt{2}}
| J,L_3\rangle &=& \sum_{L_3'} |J,L_3'\rangle D_{L_3'L_3}^J (0,\pi,\pi /2)= (-1)^{J+\frac{1}{2}L_3}|J,-L_3\rangle \,, 
\end{eqnarray}
where $D_{L_3'L_3}^J$, for given $J$, is the matrix of Wigner
functions representing the symmetry operation. 

Seeking simultaneous solutions of the above FR constraints, we obtain
the ground state as $|1,0\rangle \otimes |0,0\rangle$. The first excited
state with $I=0$ is $|3,0\rangle \otimes |0,0\rangle$, and the lowest state
with $I=1$ is $|0,0\rangle \otimes |1,0\rangle$ \cite{irwin}.

\section{Mean Charge Radius of the Quantized $B=6$ Skyrmion}
\label{sec:rad}
The state $|1,0\rangle \otimes |0,0\rangle$ of the $B=6$ Skyrmion has
the quantum numbers of the ground state of the Lithium-6 nucleus. Let us
calculate its mean charge
radius $\langle r^2 \rangle ^{\frac{1}{2}}$, defined as the square
root of
\begin{equation} 
\langle r^2 \rangle = \frac{1}{\langle\Psi | \int
  \hat{J}_0(\hbox{\bf x},t) \, d^3 x \, | \Psi \rangle}\,\langle \Psi | 
\int r^2\hat{J}_0(\hbox{\bf x},t) \, d^3 x \, | \Psi \rangle \,, 
\end{equation} 
where $\hat{J}_0$ is the electric charge density operator and $| \Psi \rangle$
represents the state with any value of $J_3$. Only the isoscalar part of
$\hat{J}_0(\hbox{\bf x},t)$ contributes to the matrix elements, and so
the integrals are pure $c$ numbers. The normalization factor is 3, the
total electric charge of Lithium-6. Therefore
\begin{equation} 
\label{eq:rad}
\langle r^2 \rangle = \frac{1}{6}\int r^2\,B_0(\hbox{\bf x}) \, d^3 x \,,
\end{equation}
using (\ref{eq:jb}). 

Using the expression (\ref{eq:baryon}) for the baryon density
in terms of the rational map ansatz, we obtain
\begin{eqnarray}
\langle r^2 \rangle &=& \frac{1}{6}\,\frac{1}{4\pi}\int 
\left( \frac{1 + |z|^2 }{ 1 + |R|^2} \left\vert\frac{dR }{ dz
}\right\vert \right)^2 \hbox{sin}\,\theta \,d\theta
\,d\phi \ \frac{2}{\pi}\int (-f')\,r^2\,\sin^2 f \, dr \,, \\
\label{eq:rad1}
&=& \frac{2}{\pi}\int (-f')\,r^2\,\sin^2 f \, dr \,.
\end{eqnarray}
The profile function $f$ can be numerically determined using a
shooting method\footnote{Thanks to Bernard M. A. G. Piette, University
of Durham, UK, for providing the C++ code}, for which the inputs are
$B$, $m$ and $\mathcal{I}$. Then the radial integral in
(\ref{eq:rad1}) can be evaluated, and the mean charge radius in Skyrme
units determined as the square root of this. $B=6$ and $\mathcal{I}=50.76$
for the rational map (\ref{eq:ratmap}), but we defer
discussion of $m$ and the actual value of $\langle r^2 \rangle
^{\frac{1}{2}}$ until section \ref{sec:reparam}.

\section{Quadrupole Moment of the Quantized $B=6$ Skyrmion}
\label{sec:quad}
An additional static electromagnetic property that is known
experimentally for the Lithium-6 nucleus to quite high precision is
its quadrupole moment. The classical quadrupole tensor of a Skyrmion
is defined as
\begin{equation}
\label{eq:q}
{Q}_{ab}=\int{(3x_ax_b - r^2\delta_{ab})\hat{J}_0(\hbox{\bf
    x},t) \, d^3x\,}
=\frac{1}{ 2}\int{(3x_ax_b - r^2\delta_{ab})B_0(\hbox{\bf x}) \,d^3x} \,,
\end{equation}
using (\ref{eq:jb}). It is
traceless by definition. The $D_{4d}$ symmetry of the $B=6$ Skyrmion 
implies that the tensor $Q_{ab}$ is diagonal, and
further that $Q_{11} = Q_{22} = -Q_{33}/2$.  

From (\ref{eq:q}) we find 
\begin{equation} 
Q_{33}=\frac{1}{2}\int{r^2
(3\cos^2 \theta -1)B_0(\hbox{\bf x}) \, d^3x} \,, 
\end{equation} 
and therefore, using the rational map ansatz,
\begin{equation}
\label{eq:q331}
Q_{33}=\frac{1}{2}\,\frac{1}{4\pi} \int (3\cos^2 \theta -1) 
\left( \frac{1 + |z|^2 }{ 1 + |R|^2}
\left\vert\frac{dR}{dz }\right\vert \right)^2 \, \hbox{sin}\,\theta \,d\theta
\,d\phi \ 
\frac{2}{\pi}\int (-f')\,r^2\,\sin^2 f \, dr \,. 
\end{equation}

For the rational map given previously for the
$B=6$ Skyrmion, 
\begin{equation*} 
\left( \frac{1 + |z|^2}{1 + |R|^2}
\left\vert\frac{dR}{dz }\right\vert \right)^2 \quad\quad\quad\quad 
\end{equation*} 
\begin{equation} 
\label{eq:long}
=4|z|^2\left(1+|z|^2\right)^2 
\left\{\frac{\left(az^8+(3a^2-1)z^4+a\right)
\left(\bar{a}\bar{z}^8+(3\bar{a}^2-1)\bar{z}^4+\bar{a}\right)
}{
\left(|a|^2|z|^{12}+|z|^8+|z|^4(1+az^4+\bar{a}\bar{z}^4)
+\bar{a}z^4+a\bar{z}^4+|a|^2\right)^2}\right\} \,, 
\end{equation} 
and substituting $z = \hbox{tan}(\theta /2)e^{i\phi}$, writing
$a=i\alpha$, where $\alpha \in \mathbb R$, and setting $\beta= 1 +
3\alpha^2$, this becomes
\begin{equation} 
\left( \frac{1 + |z|^2
}{ 1 + |R|^2} \left\vert\frac{dR}{dz }\right\vert \right)^2 = 4\tan^2
\frac{\theta}{2}\left(1+\tan^2 \frac{\theta}{2}\right)^2 \cdot $$ $$
\left\{\frac{ \alpha^2 \tan^{16} \frac{\theta}{2} + 2\alpha\beta\sin
4\phi \tan^{12} \frac{\theta}{2} +\left(\beta^2+2\alpha^2\cos
8\phi\right)\tan^8 \frac{\theta}{2} -2\alpha\beta\sin 4\phi \tan^4
\frac{\theta}{2} + \alpha^2 }{ \left(\alpha^2\tan^{12}
\frac{\theta}{2}+(1-2\alpha\sin 4\phi)\tan^8
\frac{\theta}{2}+(1+2\alpha\sin 4\phi)\tan^4
\frac{\theta}{2}+\alpha^2\right)^2}\right\} \,.
\end{equation}
A numerical integration technique can then be used to determine the
angular integral in (\ref{eq:q331}), for $\alpha = 0.16$. The result
is 
\begin{equation}
Q_{33} = 0.395\cdot \ \frac{2}{\pi}\int (-f')\,r^2\,\sin^2 f \, dr \,, 
\end{equation} 
where the radial integral is the squared charge radius (\ref{eq:rad1}). 
The accuracy of the numerical integration can be 
checked by considering the integral 
\begin{equation} 
\int \left( \frac{1 + |z|^2
}{ 1 + |R|^2} \left\vert\frac{dR}{dz }\right\vert \right)^2 
\, \hbox{sin}\,\theta \,d\theta \,d\phi \,= 4 \pi B \,.
\end{equation} 
The same procedure yields a result 75.40, which is equal to $24\pi$.

Note that $Q_{33}$ is positive, so the classical Skyrmion is prolate. 
It is helpful to consider upper and lower bounds on its value, for a 
given $B$ and $\langle r^2 \rangle$. We observe that $Q_{33}$ is 
maximal for baryon densities restricted to the $x_3$-axis and minimal 
for baryon densities restricted to the $(x_1,x_2)$-plane, hence 
\begin{equation}
-\frac{B}{2} \langle r^2 \rangle \leq Q_{33} \leq B \langle r^2
 \rangle \,.
\end{equation}
For the $B=6$ case,
\begin{equation}
\label{eq:lowerupper}
-3 \leq Q_{33}/\langle r^2 \rangle \leq 6 \,.
\end{equation}
Within this range, the actual value $Q_{33}/\langle r^2 \rangle = 0.395$ 
is rather close to zero.

In the quantum state the Skyrmion occurs in all possible
orientations. Substituting 
\begin{equation}
\hat{U}(\hbox{\bf x})=A_1 U_0(D(A_2)\hbox{\bf x})A_1^{\dag} \,,
\end{equation} 
we obtain 
\begin{equation}
\hat{Q}_{ij} = D(A_2)_{ia}^T Q_{ab} D(A_2)_{bj} \,, 
\end{equation} 
for the rotated classical Skyrmion. 
We observe that $\hat{Q}_{ij}$ is independent of $A_1$. 
The symmetry relations reduce this quadrupole moment operator to the expression
\begin{equation} 
\hat{Q}_{ij} = \frac{1}{2}Q_{33}\left(3D(A_2)_{i3}^TD(A_2)_{3j} -
\delta_{ij}\right) \,,
\end{equation} 
and thus 
\begin{equation} 
\hat{Q}_{33} =
\frac{1}{2}Q_{33}\left(3\cos^2 \theta -
  1\right)=Q_{33}D_{00}^2(\phi,\theta,\psi) \,, 
\end{equation} 
where $\phi, \theta, \psi$ now denote the Euler angles parametrising
$A_2$, and $D_{00}^2 (\phi,\theta,\psi)$ is a Wigner function.

The wavefunction $\Psi$ is the product of Wigner functions
$D_{sm}^j(\phi,\theta,\psi)\otimes D_{tn}^i(\alpha,\beta,\gamma)$ on
$SO(3)^J \times SO(3)^I$. For the ground state
of the $B=6$ Skyrmion, with $J=1$ and $I=0$, and $J_3 = m$,
\begin{equation} 
\Psi = \sqrt\frac{3}{ {8\pi^2}}D_{0m}^1(\phi,\theta,\psi) \,. 
\end{equation}
The quadrupole moment of the quantized $B=6$ Skyrmion is defined as 
the expectation value of $\hat{Q}_{33}$ in the state with 
$J_3=1$ (by convention the quadrupole moment is measured in the 
top spin state), i.e. 
\begin{equation}
Q=\langle\Psi_{J_3=1} | \hat{Q}_{33} | \Psi_{J_3=1}\rangle 
= \frac{3}{8\pi^2} Q_{33} \int
  D_{00}^2\,|D_{01}^1|^2\,\hbox{sin}\,\theta \, d\phi \, d\theta \, d\psi \,,
\end{equation} 
and using the well-known expression in terms of Wigner $3j$-symbols 
for the integral of the product of three Wigner functions, 
\begin{equation} 
\int D_{ab}^j\, D_{cd}^{j'} \,D_{ef}^{j''}
\,\hbox{sin}\,\theta \,d\phi \,d\theta \,d\psi = 8\pi^2 \left(
\begin{array}{ccc} j & j' & j'' \\ a & c & e \end{array} \right) \left(
\begin{array}{ccc} j & j' & j'' \\ b & d & f \end{array} \right) \,, 
\end{equation} 
we find that 
\begin{eqnarray} 
Q&=&-\frac{3}{8\pi^2} Q_{33} \int D_{00}^2
\,D_{01}^1\,D_{0\,-1}^1\,\hbox{sin}\,\theta \,d\theta
\,d\phi\,d\psi,\,\,\,\, \hbox{as}\,\,\,\,
{D_{m'm}^j}^* = (-1)^{m'-m}D_{-m'-m}^j \nonumber\\ &=&-3Q_{33}\left(
\begin{array}{ccc} 2 & 1 & 1 \\ 0 & 0 & 0 \end{array} \right) \left(
\begin{array}{ccc} 2 & 1 & 1 \\ 0 & 1 & -1 \end{array} \right) \,, \nonumber\\
&=&-3Q_{33}\sqrt{\frac{2}{15}}\sqrt{\frac{1}{30}} \,, \nonumber\\\
\label{eq:fifth} &=&-\frac{1}{5}Q_{33} \,.
\end{eqnarray}

This is the standard result for an axisymmetric system in a $J=1$
state. Note that a prolate classical shape ($Q_{33}$ positive) gives
an oblate quadrupole moment ($Q$ negative), and vice versa, because
the prolate object can be regarded as spinning about its short axis
aligned along the third axis in space when $J_3=1$. From 
(\ref{eq:lowerupper}) we deduce the inequalities
\begin{equation}
\label{eq:lowerupperQ}
-1.2 \leq Q/\langle r^2 \rangle \leq 0.6 \,.
\end{equation} 

We note that to obtain physically meaningful results for both the mean
charge radius and the quadrupole moment, it is necessary to convert
back to physical units, by reintroducing the Skyrme model parameters,
and using the conversion factor $\hbar c =
197.3\,\hbox{MeV}\,\hbox{fm}$. We do this in the next section.

\section{Reparametrising the Skyrme model}
\label{sec:reparam}
The parameters $e$ and $F_\pi$ can be fixed in a number of ways. It
has been common practice to use the set of parameters given in
\cite{an}, specifically 
\begin{equation}
\label{eq:prov}
e=4.84,\,\,\,F_\pi = 108\,\hbox{MeV}\,\,\,
\hbox{and}\,\,\,m_\pi=138\,\hbox{MeV}\,\,\, 
(\hbox{which implies}~m=0.528) 
\end{equation}
for the Skyrme model with the physical pion mass taken into
account. In \cite{an}, the values of $e$ and $F_\pi$ were tuned to
reproduce the masses of the proton and the delta resonance. This
parameter set was adjusted to optimise the predictions of the model in
the $B=1$ sector at the expense of the $B=0$ sector, which requires
$F_\pi = 186\,\hbox{MeV}$. It is therefore not the
optimal parameter set with which to describe the higher baryon number
sectors of the Skyrme model. Nevertheless, in \cite{bc} this parameter 
set was used in a quantization of the collective coordinates of 
the toroidal $B=2$ Skyrmion. The resulting quantum state is too small
compared to the size of the deuteron, and it is too tightly bound. 
However, this discrepancy was
dealt with by the quantization of selected vibrational modes of the
$B=2$ Skyrmion in \cite{lms}. We therefore do not propose to change
$e$ and $F_\pi$ for $B=1$ or $B=2$. However, for higher $B$, it
becomes technically very difficult to perform a quantization of the
possible vibrational modes. Instead, we consider rescaling the
parameters in order to fit the Skyrme model to experimental nuclear data,
whilst suppressing the quantization of vibrational modes. This is a
type of renormalization.

We have first determined $\langle r^2 \rangle ^{\frac{1}{2}}$ and $Q$
for the $B=6$ Skyrmion using the set of parameters given in \cite{an}, 
and setting $a=0.16i$ (the value of $a$ that minimises $\mathcal{I}$, 
as required by the rational map ansatz). After evaluating the radial
integral (\ref{eq:rad1}), we find
\begin{equation}
\langle r^2 \rangle ^{\frac{1}{2}} = 1.48\,\hbox{fm}\,,\,\,\,
Q = -0.173\,(\hbox{fm})^2 \,.
\end{equation}

In \cite{dejager} the experimental root mean square radius of the 
charge density distribution of the Lithium-6 nucleus is given 
as $\langle r^2 \rangle ^{\frac{1}{2}} = 2.55\,\hbox{fm}$,
substantially larger than $1.48\,\hbox{fm}$, and its quadrupole moment is given
as $-0.82 \times 10^{-3}\, \hbox{barns} = -0.082\,(\hbox{fm})^2$ in 
\cite{lithium}. We can
change the length scale $2/eF_\pi$, to obtain a mean charge radius in
close agreement with experiment. This results in the
dimensionless mass $m=2m_\pi/eF_\pi$, with $m_\pi = 138\,\hbox{MeV}$
as before, being changed, and so the profile function and hence the
radial integral in the expressions for the mean charge radius and
quadrupole moment is also modified. By iteration, it is found
that by setting $m=1.125$, and hence $eF_\pi = 245.34\,\hbox{MeV}$, we
obtain $\langle r^2 \rangle ^{\frac{1}{2}} = 2.55\,\hbox{fm}$, which
fits the data.

With this choice for $eF_\pi$, the theoretical prediction of the
quadrupole moment becomes $Q = -0.512\,(\hbox{fm})^2$. Although the
sign is right, this is unfortunately
about six times the experimental value. One must bear
in mind that the rational map ansatz does not generally determine
exact solutions of the Skyrme field equation, but rather
approximations to solutions. It is useful to allow a slight
modification of the rational map (\ref{eq:ratmap}), while preserving
its symmetry and prolateness. If we take $a = 0.1933i$, rather than $0.16i$, 
and use the new value for $eF_\pi$, we obtain a quadrupole
moment in very close agreement with experiment: $Q = -0.082\,(\hbox{fm})^2$.

We note that this modification of the rational map leads to the value of
the integral $\mathcal{I}$ being slightly modified (from 50.76 to
51.49, i.e. not quite the minimum as required by the rational map
ansatz), which subsequently leads to the profile function and hence the
radial integral in both the expressions for the mean charge radius and
the quadrupole moment being modified, but only slightly. In
summary, using $eF_\pi = 245.34\,\hbox{MeV}$, $m_\pi = 138\,\hbox{MeV}$ 
(which implies $m=1.125$), and $a=0.1933i$, we find 
\begin{equation}
\langle r^2 \rangle ^{\frac{1}{2}} = 2.55\,\hbox{fm} \,,\,\,\,
Q = -0.082\,(\hbox{fm})^2 \,,
\end{equation}
and hence $Q/\langle r^2 \rangle = -0.0127$.
By virtue of (\ref{eq:lowerupperQ}), we consider the quadrupole moment
of the Lithium-6 nucleus to be only slightly oblate. The change we
needed to make to the parameter $a$ should not be regarded as a
large change.

It remains to consider the energy scale $F_\pi/4e$ of the Skyrme
model. Assuming negligible spin kinetic energy (it is expected to
be in the region of $1\,\hbox{MeV}$), we may identify the static
Skyrmion energy (\ref{eq:e}) with the mass of the Lithium-6 nucleus,
$M$, given by
\begin{equation}
M = 6m_N - E_{binding} = 5601\,\hbox{MeV},
\end{equation}
where $m_N = 939\,\hbox{MeV}$ is the average mass of a nucleon and
$E_{binding}=32\,\hbox{MeV}$ is the total binding energy of the
nucleus \cite{binding}.
The static Skyrmion energy is determined numerically, and we find
for $m=1.125$, after converting to physical units, that
\begin{equation}
E = \frac{F_\pi}{4e}\cdot 972 \,.
\end{equation}
Equating this to the mass of the nucleus implies that
\begin{equation}
F_\pi / 4e = 5.76\,\hbox{MeV} \,,
\end{equation}
and combined with $eF_\pi = 245.34\,\hbox{MeV}$ we obtain, finally
\begin{equation}
\label{eq:newparam}
e=3.263,\,\,\,F_\pi = 75.20\,\hbox{MeV}\,\,\, 
\hbox{and}\,\,\,m_\pi=138\,\hbox{MeV}\,\,\, (\hbox{which
  implies}~m=1.125) \,. 
\end{equation}

We conclude that with this choice of parameters, and taking
$a=0.1933i$, the mass and static electromagnetic properties of the quantized 
$B=6$ Skyrmion are in close agreement with those experimentally determined
for the Lithium-6 nucleus. As mentioned previously, this value of
$F_\pi$ is considered to be a renormalized pion decay constant, and
the fact that it is so much less than the experimental value is
therefore not necessarily conflicting.

\section{Reconsidering the $B=2$ and $B=4$ Skyrmions}
\label{sec:b=2}
\subsection{The quantized toroidal $B=2$ Skyrmion}
The minimal energy solution in the $B=2$ sector is axially symmetric,
and has the shape of a torus. In \cite{bc}, Braaten and Carson performed the collective coordinate quantization of the solution (see also \cite{kop}). Using the same notation as that
used above for the $B=6$ case, it was found that the axial symmetry of
the solution requires
\begin{equation}
(2K_3 + L_3)|\Psi\rangle = 0.
\end{equation}
In addition to this constraint, the FR constraint associated with the
discrete rotations by $\pi$ in the $D_2$ symmetry group is
\begin{equation}
e^{i\pi K_1}e^{i\pi L_1}|\Psi\rangle = -|\Psi\rangle.
\end{equation}
The ground state is the $J=1$, $I=0$ state 
$|1,0\rangle \otimes |0,0\rangle$, which has the quantum numbers of
the deuteron. The first excited state $|0,0\rangle \otimes |
1,0\rangle$ may be identified with the isovector $^1 S_0$ state of the
two-nucleon system.

Further properties of the quantized Skyrmion were also
determined, taking the Skyrme model parameters to be as in
(\ref{eq:prov}), and compared to experimental data on the deuteron.
An expression equivalent to (\ref{eq:rad}) is obtained for the mean
charge radius of the $B=2$ Skyrmion. The expression for the quantum
quadrupole moment in this case is also given by
$Q=-\frac{1}{5}Q_{33}$, where $Q_{33}$ is negative here, because the
classical Skyrmion is oblate. The quantity $Q_{33}$ was evaluated
in \cite{bc} using the numerical, exact classical solution 
and its corresponding baryon number density. The predictions were:
\begin{eqnarray}
\label{eq:bcpred1}
\langle r^2 \rangle ^{\frac{1}{2}} &=& 0.92\,\hbox{fm}, \\
\label{eq:bcpred2}
Q &=& 0.082\,({\hbox{fm}})^2 \,, \\
\label{eq:bcpred3}
M &=& 1720\,\hbox{MeV}\,,
\end{eqnarray}
compared to experimental data of $\langle r^2 \rangle ^{\frac{1}{2}} =
2.095\,\hbox{fm}$ \cite{ericson}, 
$Q= 0.2859\,({\hbox{fm}})^2$ \cite{deut} and $M =
1876\,\hbox{MeV}$. The mass of the nucleus (\ref{eq:bcpred3}) is given by the sum of the classical mass, equal to $1659\,\hbox{MeV}$, and a spin correction equal to $1/V_{11} = 61\,\hbox{MeV}$. It was suggested in \cite{bc} that these discrepancies could be an
artifact of the Skyrme model parameters, and that the disagreement
with experiment could be assuaged by an adjustment of $e$ and $F_\pi$
such that $eF_\pi$ is approximately halved and $F_\pi/e$ is unchanged. The new
parameter set which we propose is close to this, but we also take into
account the corresponding change of $m$ when fitting to experimental 
data.

\subsection{Attractive channel Skyrmions and the deuteron}
Because the deuteron appears too small and too tightly bound according
to \cite{bc}, it was suggested in \cite{lms} that the deuteron is
better described as a quantum state on a
ten-dimensional manifold of $B=2$ Skyrme fields. This manifold
includes the toroidal configurations of minimal energy (i.e. the
eight-dimensional manifold considered in \cite{bc}), and deformations
of these into configurations which are approximately the product 
of two separated $B=1$ Skyrmions in
the most attractive relative orientation. The ten coordinates of this
manifold are the separation parameter, overall translations, rotations
and isorotations. A unique bound quantum state is found with the
quantum numbers of the deuteron. The values of
$e$ and $F_\pi$ were taken to be 4.84 and 108$\,\hbox{MeV}$
respectively, i.e. those of \cite{an} and \cite{bc}, but for technical 
reasons the pion mass was taken to be zero. In this model, the predictions are
\begin{eqnarray}
\langle r^2 \rangle ^{\frac{1}{2}} &=& 2.18\,\hbox{fm},\\
Q &=& 0.83\,\hbox{(fm)}^2,\\
\label{eq:1872}
M &=& 1872\,\hbox{MeV}.
\end{eqnarray}
The calculated deuteron binding energy of $6\,\hbox{MeV}$ is very close to the experimentally determined binding energy of the deuteron, but the total mass could not be straightforwardly determined. The prediction of the deuteron mass (\ref{eq:1872}) needs to be interpreted as that of the sum of the proton and neutron masses minus this binding energy. The prediction for the mean charge radius is in very close agreement
with experiment. However, the deuteron's quadrupole moment has become
much larger than the experimental value.
This is explained in \cite{lms} as follows: The tensor force
is responsible for the existence of a d-wave contribution to the
deuteron wave function. In the absence of a d-wave, $Q$ would be
zero. The size of $Q$ is therefore a measure of the d-wave
probability, which in turn indicates the strength of the tensor
force. The truncation to the space of attractive channel fields
systematically overestimates the strength of the tensor
force. Therefore it is not surprising that the theoretical prediction
for $Q$ is rather large.

\subsection{Is reparametrisation helpful in the $B=2$ sector?}
Let us reconsider the collective coordinate quantization of \cite{bc},
ignoring vibrational modes, but using our new parameter set
(\ref{eq:newparam}). To avoid recalculating the exact $B=2$ Skyrmion 
solution for various values of the pion mass parameter $m$, we use the rational map ansatz. 
There is a unique map here, namely $R(z)=z^2$ \cite{houghton}. 
This gives, for the quantized $B=2$ Skyrmion,
\begin{equation}
\langle r^2 \rangle 
= \frac{1}{2}\,\frac{1}{4\pi}\int \left( \frac{1 + |z|^2 }{ 1 + |R|^2} 
\left\vert\frac{dR}{dz}\right\vert \right)^2 \hbox{sin}\,\theta \,d\theta
\,d\phi \ \frac{2}{\pi} \int (-f')\,r^2\,\sin^2 f \, dr \, 
= \frac{2}{\pi}\int (-f')\,r^2\,\sin^2 f \, dr \,, 
\end{equation}
and
\begin{equation}
Q = -\frac{1}{5}\,\frac{1}{2}\,\frac{1}{4\pi} \int 
(3\cos^2 \theta -1) \left( \frac{1 + |z|^2}{1 + |R|^2}
\left\vert\frac{dR}{dz }\right\vert \right)^2 \, \hbox{sin}\,\theta \,d\theta
\,d\phi \ \frac{2}{\pi}\int (-f')\,r^2\,\sin^2 f \, dr \,. 
\end{equation}
It is straightforward to calculate the profile function $f$ in this case,
and to perform the numerical integrations.
Converting back to physical units, and using first the traditional values of 
$e$, $F_\pi$ and $m$ given by (\ref{eq:prov}), we obtain
\begin{eqnarray}
\langle r^2 \rangle ^{\frac{1}{2}} &=& 0.940\,\hbox{fm},\\
Q &=& 0.102\, \hbox{(fm)}^2,\\
\label{eq:1757}
M &=& 1757\,\hbox{MeV}.
\end{eqnarray}
which are reasonably close to the values
(\ref{eq:bcpred1})--(\ref{eq:bcpred3}) predicted in \cite{bc}. We have again assumed that the mass of the nucleus (\ref{eq:1757}) can be equated to the sum of the static Skyrmion energy ($1696\,\hbox{MeV}$) and the spin correction which we take to be the same as in \cite{bc} ($61\,\hbox{MeV}$). Comparing this value of the static Skyrmion energy to the classical mass given in \cite{bc}, we see that the rational map ansatz gives a very good approximation.

Using the new parameters (\ref{eq:newparam}), and their associated 
length and energy scales, we find after recalculating the radial 
integral (which we recall is a function of $m$)
\begin{eqnarray}
\label{eq:1644}
\langle r^2 \rangle ^{\frac{1}{2}} &=& 1.644\,\hbox{fm},\\
Q &=& 0.311\, \hbox{(fm)}^2,\\
\label{eq:1969}
M &=& 1969\,\hbox{MeV}.
\end{eqnarray}
The deuteron mass (\ref{eq:1969}) is again assumed to be equal to the static Skyrmion energy ($1950\,\hbox{MeV}$) plus an estimated spin correction equal to $61\,\hbox{MeV}$ times the ratio of the squares of the predicted mean charge radii (\ref{eq:bcpred1}) and (\ref{eq:1644}), which allows for the increase in the moment of inertia. Comparing these with the experimental values, we conclude that the new parameters give quite a good description of the deuteron, better than
those in \cite{bc}, and provide 
an alternative to taking account of the soft vibrations. 

\subsection{The $B=4$ Skyrmion and the $\alpha$-particle}
The $B=4$ Skyrmion is cubically symmetric, and was first quantized by Walhout \cite{walhout}, using the exact classical solution and the traditional parameter set. The rational map 
ansatz for this Skyrmion uses the 
unique map of degree 4 with cubic symmetry \cite{houghton},
\begin{equation}
R(z) = \frac{z^4+2\sqrt{3}iz^2 +1}{z^4-2\sqrt{3}iz^2 +1} \,.
\end{equation}
The FR constraints corresponding to the cubic symmetry 
were determined in \cite{irwin} as
\begin{eqnarray}
e^{\frac{2\pi i}{3\sqrt{3}}(L_1+L_2+L_3)} 
e^{\frac{2\pi i}{3\sqrt{3}}(K_1+K_2+K_3)}|\Psi \rangle &=& |\Psi \rangle, \\
e^{i\frac{\pi}{2}L_3} e^{i\frac{\pi}{\sqrt{2}}(K_1-K_2)}|\Psi \rangle 
&=& |\Psi \rangle \,.
\end{eqnarray}
The ground state, in the basis introduced previously,
is $|0,0\rangle \otimes |0,0\rangle$, which trivially satisfies the
constraints. This state may be identified with the $\alpha$-particle.

If vibrational modes are ignored, then the traditional parameters give
a much too small and tightly bound $\alpha$-particle. 
Walhout's rather complicated analysis
takes into account a number of the vibrational modes of the $B=4$ 
Skyrmion, obtaining $\langle r^2 \rangle ^{\frac{1}{2}} = 1.58\,\hbox{fm}$
and $M = 3677\,\hbox{MeV}$, compared with the experimental values 
$\langle r^2 \rangle ^{\frac{1}{2}} = 1.71\,\hbox{fm}$ \cite{dejager} 
and $M = 3727\,\hbox{MeV}$.

We wish to ignore the vibrational modes, but in compensation, use 
our new parameter set (\ref{eq:newparam}). As the
spin of the $B=4$ Skyrmion is zero, there is no quadrupole moment to 
determine, just the mean charge radius. The mass of the
quantized Skyrmion is equal to the classical mass, as
there is no spin energy contribution.
Numerically computing the profile function (with inputs ${\cal{I}} =
20.65$ and $m=1.125$), and numerically integrating the corresponding
densities, we find
\begin{eqnarray}
\langle r^2 \rangle ^{\frac{1}{2}} &=& 2.13\,\hbox{fm}, \\
M &=& 3679\,\hbox{MeV} \,,
\end{eqnarray}
which agrees reasonably well with experiment. 

\section{Conclusion}
\label{sec:conc}

We conclude that in order to accurately predict certain properties of
nuclei (with baryon numbers greater than two) using the Skyrme model,
it is necessary to rescale the Skyrme model parameters, if the
quantization of vibrational modes is not going to be considered in the
calculation. We have in this paper explicitly determined the values of
these parameters in the $B=6$ case, and discussed how well they work
in the $B=2$ and $B=4$ cases. It would be interesting to see
how well they work for Skyrmions with other baryon numbers.

Our suggestion that the dimensionless pion mass $m$ should be increased 
from 0.528 to 1.125 backs up the suggestion made in \cite{bs2}, in which 
it was found that as the pion mass is increased its effect becomes important
for Skyrmions of all baryon numbers. In particular, the hollow,
polyhedral shell-like solutions which exist for a large range of
baryon numbers up to 20 or 30, and are stable for zero pion mass, 
become unstable for baryon number eight and above when $m$ is of order
unity. Instead, the stable solutions become more solid structures,
some of which are related to chunks of the Skyrme crystal. 
This improves the qualitative fit of the Skyrme model to
real nuclei. Also, in \cite{bms}, low energy
Skyrmion solutions composed of charge four sub-units were found for
baryon numbers a multiple of four, with $m=1$. This
is the Skyrmion analogue of the $\alpha$-particle model of nuclei.

The standard values of the Skyrme parameters originate from \cite{an},
in which attention was restricted to the single nucleon. In recent
years, the substantial increase in results available for Skyrmions
over a range of baryon numbers enables us to fit the Skyrme parameters
to experimental data for larger nuclei, with a view to modelling 
nuclei of all baryon numbers. The parameters we propose here should be
regarded as provisional, since they rely on just a few basic
properties of Lithium-6, and depend on the rational map 
approximation. More work remains to be done within the Skyrme model to 
see if one set of Skyrme parameters can fit experimental 
data over a range of baryon numbers. Both static
properties of nuclei, like their masses and electric charge distribution, 
and also the excitation energies of higher spin states, need to be
considered further.

\subsection*{Acknowledgements}
NSM thanks Vladimir B. Kopeliovich for correspondence concerning the
Skyrme model parameters. SWW thanks Bernard M. A. G. Piette for
providing the C++ code with which the rational map profile functions
were numerically determined, and thanks Steffen Krusch for helpful
discussions. SWW would like to gratefully acknowledge funding from PPARC.


\begin{thebibliography}{99}

\bibitem{an} G. S. Adkins and C. R. Nappi,
\emph{The Skyrme model with pion masses},
Nucl. Phys. B233: 109 (1984)

\bibitem{bms} R. A. Battye, N. S. Manton and P. M. Sutcliffe,
\emph{Skyrmions and the $\alpha$-particle model of nuclei},
hep-th/0605284 (2006) (to appear in Proc. Roy. Soc. A)

\bibitem{bs} R. A. Battye and P. M. Sutcliffe,
\emph{Symmetric Skyrmions},
Phys. Rev. Lett. 79: 363 (1997)

\bibitem{bs1} R. A. Battye and P. M. Sutcliffe,
\emph{Skyrmions, fullerenes and rational maps},
Rev. Math. Phys. 14: 29 (2002)

\bibitem{bs2} R. A. Battye and P. M. Sutcliffe,
\emph{Skyrmions and the pion mass},
Nucl. Phys. B705: 384 (2005)

\bibitem{bc} E. Braaten and L. Carson,
\emph{Deuteron as a toroidal Skyrmion},
Phys. Rev. D38: 3525 (1988)

\bibitem{bc1} E. Braaten and L. Carson,
\emph{Deuteron as a toroidal Skyrmion: Electromagnetic form factors}
Phys. Rev. D39: 838 (1989)

\bibitem{callanwitten} C. G. Callan and E. Witten,
\emph{Monopole catalysis of Skyrmion decay},
Nucl. Phys. B239: 161 (1984)

\bibitem{lithium} J. Cederberg et al., \emph{Nuclear
electric quadrupole moment of $^6 \hbox{Li}$}, Phys. Rev. A57: 2539 (1998)

\bibitem{dejager} C. W. De Jager, H. De Vries and C. De Vries,
\emph{Nuclear charge- and magnetization-density-distribution parameters from
elastic electron scattering},
Atom. Data. Nucl. Data Tab. 14: 479 (1974)

\bibitem{ericson} T. E. O. Ericson,
\emph{The deuteron properties},
Nucl. Phys. A416: 281 (1984)

\bibitem{fr} D. Finkelstein and J. Rubinstein,
\emph{Connection between spin, statistics and kinks},
J. Math. Phys. 9: 1762 (1968)

\bibitem{hm} C. Houghton and S. Magee,
\emph{The effect of pion mass on Skyrme configurations},
hep-th/0602227 (2006)

\bibitem{houghton} C. J. Houghton, N. S. Manton and P. M. Sutcliffe,
\emph{Rational maps, monopoles and Skyrmions},
Nucl. Phys. B510: 507 (1998)

\bibitem{irwin} P. Irwin,
\emph{Zero mode quantization of multi-Skyrmions},
Phys. Rev. D61: 114024 (2000)

\bibitem{kop} V. B. Kopeliovich,
\emph{Quantization of the axially-symmetric systems' rotations in the Skyrme model (in Russian)},
Yad. Fiz. 47: 1495 (1988)

\bibitem{krusch} S. Krusch,
\emph{Homotopy of rational maps and the quantization of Skyrmions},
Ann. Phys. 304: 103 (2003)
 
\bibitem{landau} L. D. Landau and E. M. Lifschitz,
\emph{Quantum Mechanics - Course of Theoretical Physics Vol. 3, 3rd Edition},
Butterworth-Heinemann (1977)

\bibitem{lms} R. A. Leese, N. S. Manton and B. J. Schroers,
\emph{Attractive channel Skyrmions and the deuteron},
Nucl. Phys. B442: 228 (1995)

\bibitem{manko} O. V. Manko and N. S. Manton,
\emph{Angularly localized Skyrmions},
J. Phys. A: Math. Gen. 39: 1507 (2006)

\bibitem{mantonpiette} N. S. Manton and B. M. A. G. Piette,
\emph{Understanding Skyrmions using rational maps}, Prog.
Math. 201: 469 (2001)

\bibitem{manton} N. S. Manton and P. M. Sutcliffe,
\emph{Topological Solitons},
Cambridge University Press (2004)

\bibitem{binding} K. Saito et al.,
\emph{Structure functions of unstable lithium isotopes},
Nucl. Phys. A705: 119 (2002)

\bibitem{deut}  V. S. Shirley, C. M. Baglin, S. Y. F. Chu and J. Zipkin Eds. 
\emph{Table of Isotopes, Vol.1, 8th Edition},
Wiley, New York (1996)

\bibitem{skyrme} T. H. R. Skyrme,
\emph{A nonlinear field theory},
Proc. Roy. Soc. A260: 127 (1961)

\bibitem{walhout} T. S. Walhout,
\emph{Quantizing the four-baryon Skyrmion},
Nucl. Phys. A547: 423 (1992)

\bibitem{wittenglobal} E. Witten,
\emph{Global aspects of current algebra},
Nucl. Phys. B223: 422 (1983)

\bibitem{wittencurrent} E. Witten,
\emph{Current algebra, baryons, and quark confinement},
Nucl. Phys. B223: 433 (1983)

\end{thebibliography}
\end{document}